\documentclass[journal=apchd5,manuscript=article,layout=twocolumn]{achemso}

\usepackage[T1]{fontenc}
\usepackage{amsmath}
\usepackage{graphicx}
\usepackage[separate-uncertainty = true,multi-part-units=single]{siunitx}

\newcommand*{\wExNeutral}{1.951} 
\newcommand*{\wExTrion}{1.922} 
\newcommand*{\wEx}{1.963} 
\newcommand*{\gammaEx}{28} 

\newcommand*{\gammaSp}{170} 

\newcommand*{\coupling}{64} 
\newcommand*{\couplingDelta}{3} 
\newcommand*{\rabi}{108} 
\newcommand*{\rabiDelta}{8} 
\newcommand*{\gammaAvg}{99} 
\newcommand*{\gammaAvgDelta}{3} 

\newcommand*{\readoffDelta}{5} 

\newcommand*{\QMax}{13.5} 

\newcommand*{\wExMulti}{1.942} 
\newcommand*{\gammaExMulti}{44} 
\newcommand*{\couplingMulti}{93} 
\newcommand*{\couplingMultiDelta}{4} 
\newcommand*{\rabiMulti}{175} 
\newcommand*{\rabiMultiDelta}{9} 
\newcommand*{\gIncrease}{45} 
\newcommand*{\rabiIncrease}{62} 

\newcommand*{\wsHeight}{0.87} 

\newcommand*{\rOne}{19}
\newcommand*{\rTwo}{23}
\newcommand*{\rThree}{25}
\newcommand*{\rFour}{27}
\newcommand*{\rFive}{29}

\newcommand*{\suppPL}{1} 
\newcommand*{\suppEnsemble}{2} 
\newcommand*{\suppReflFit}{3} 
\newcommand*{\suppComSpec}{4}
\newcommand*{\suppAvoidCrossCom}{5}
\newcommand*{\suppNorJC}{6}

\newcommand*{\suppReflMonoPar}{1}
\newcommand*{\suppReflMultiPar}{2}

\author{Mathias Geisler}
\affiliation{Department of Photonics Engineering, Technical University of Denmark, DK-2800 Kgs.~Lyngby, Denmark}
\alsoaffiliation{Center for Nanostructured Graphene (CNG), Technical University of Denmark, DK-2800 Kgs.~Lyngby, Denmark}

\author{Ximin Cui}
\affiliation{Department of Physics, The Chinese University of Hong Kong, Shatin, Hong Kong SAR, China}

\author{Jianfang Wang}
\affiliation{Department of Physics, The Chinese University of Hong Kong, Shatin, Hong Kong SAR, China}

\author{Tomas~Rindzevicius}
\affiliation{Department of Micro- and Nanotechnology, Technical University of Denmark, DK-2800 Kgs.~Lyngby, Denmark}
\alsoaffiliation{DNRF and Villum Fonden Center for Intelligent Drug Delivery and Sensing Using Microcontainers and Nanomechanics (IDUN), Technical University of Denmark, DK-2800 Kgs.~Lyngby, Denmark}

\author{Lene~Gammelgaard}
\affiliation{Center for Nanostructured Graphene (CNG), Technical University of Denmark, DK-2800 Kgs.~Lyngby, Denmark}
\alsoaffiliation{Department of Micro- and Nanotechnology, Technical University of Denmark, DK-2800 Kgs.~Lyngby, Denmark}

\author{Bjarke S. Jessen}
\affiliation{Center for Nanostructured Graphene (CNG), Technical University of Denmark, DK-2800 Kgs.~Lyngby, Denmark}
\alsoaffiliation{Department of Micro- and Nanotechnology, Technical University of Denmark, DK-2800 Kgs.~Lyngby, Denmark}

\author{P.~A.~D. Gon\c{c}alves}
\affiliation{Department of Photonics Engineering, Technical University of Denmark, DK-2800 Kgs.~Lyngby, Denmark}
\alsoaffiliation{Center for Nanostructured Graphene (CNG), Technical University of Denmark, DK-2800 Kgs.~Lyngby, Denmark}
\alsoaffiliation{Center for Nano Optics, University of Southern Denmark, DK-5230 Odense M, Denmark}

\author{Francesco~Todisco}
\affiliation{Center for Nano Optics, University of Southern Denmark, DK-5230 Odense M, Denmark}

\author{Peter~B{\o}ggild}
\affiliation{Center for Nanostructured Graphene (CNG), Technical University of Denmark, DK-2800 Kgs.~Lyngby, Denmark}
\alsoaffiliation{Department of Micro- and Nanotechnology, Technical University of Denmark, DK-2800 Kgs.~Lyngby, Denmark}

\author{Anja~Boisen}
\affiliation{Department of Micro- and Nanotechnology, Technical University of Denmark, DK-2800 Kgs.~Lyngby, Denmark}
\alsoaffiliation{DNRF and Villum Fonden Center for Intelligent Drug Delivery and Sensing Using Microcontainers and Nanomechanics (IDUN), Technical University of Denmark, DK-2800 Kgs.~Lyngby, Denmark}

\author{Martijn Wubs}
\affiliation{Department of Photonics Engineering, Technical University of Denmark, DK-2800 Kgs.~Lyngby, Denmark}
\alsoaffiliation{Center for Nanostructured Graphene (CNG), Technical University of Denmark, DK-2800 Kgs.~Lyngby, Denmark}

\author{N.~Asger~Mortensen}
\affiliation{Center for Nanostructured Graphene (CNG), Technical University of Denmark, DK-2800 Kgs.~Lyngby, Denmark}
\alsoaffiliation{Center for Nano Optics, University of Southern Denmark, DK-5230 Odense M, Denmark}
\alsoaffiliation{Danish Institute for Advanced Study, University of Southern Denmark, DK-5230 Odense~M, Denmark}

\author{Sanshui Xiao}
\affiliation{Department of Photonics Engineering, Technical University of Denmark, DK-2800 Kgs.~Lyngby, Denmark}
\alsoaffiliation{Center for Nanostructured Graphene (CNG), Technical University of Denmark, DK-2800 Kgs.~Lyngby, Denmark}

\author{Nicolas Stenger}
\affiliation{Department of Photonics Engineering, Technical University of Denmark, DK-2800 Kgs.~Lyngby, Denmark}
\alsoaffiliation{Center for Nanostructured Graphene (CNG), Technical University of Denmark, DK-2800 Kgs.~Lyngby, Denmark}
\email{niste@fotonik.dtu.dk}

\title{Single-crystalline gold nanodisks on WS\(_{2}\) mono- and multilayers: Strong coupling at room temperature}


\begin{document}

\begin{abstract}
Engineering light-matter interactions up to the strong-coupling regime at room temperature is one of the cornerstones of modern nanophotonics. Achieving this goal will enable new platforms for potential applications such as quantum information processing, quantum light sources and even quantum metrology. Materials like transition metal dichalcogenides (TMDC) and in particular tungsten disulfide (WS\(_{2}\)) possess large transition dipole moments comparable to semiconductor-based quantum dots, and strong exciton binding energies allowing the detailed exploration of light-matter interactions at room temperature. Additionally, recent works have shown that coupling TMDCs to plasmonic nanocavities with light tightly focused on the nanometer scale can reach the strong-coupling regime at ambient conditions. Here, we use ultra-thin single-crystalline gold nanodisks featuring large in-plane electromagnetic dipole moments aligned with the exciton transition-dipole moments located in monolayer WS\(_{2}\). Through scattering and reflection spectroscopy we demonstrate strong coupling at room temperature with a Rabi splitting of $\sim$\SI{\rabi}{meV}. In order to go further into the strong-coupling regime and inspired by recent experimental work by St{\"u}hrenberg et al., we couple these nanodisks to multilayer WS\(_{2}\). Due to an increase in the number of excitons coupled to our nanodisks, we achieve a Rabi splitting of $\sim$\SI{\rabiMulti}{meV}, a major increase of \SI{\rabiIncrease}{\%}. To our knowledge, this is the highest Rabi splitting reported for TMDCs coupled to open plasmonic cavities. Our results suggest that ultra-thin single-crystalline gold nanodisks coupled to WS\(_{2}\) represent an exquisite platform to explore light-matter interactions.
\end{abstract}

Efficient coupling between light and matter is important for many applications as well as for fundamental research within the field of nanophotonics. When the exchange rate~of energy between an emitter and a cavity exceeds their intrinsic dephasing rates, the system enters the strong-coupling regime where new hybrid eigenstates form that are part light and part matter~\cite{Yoshie2004}. Reaching the strong-coupling limit in different systems has among others allowed for the study of quantum electrodynamics~\cite{Wallraff2004}, experimental realization of low-threshold lasers~\cite{McKeever2003}, and for coupling a single quantum of energy to nitrogen-vacancy centers in diamond~\cite{Marcos2010,Zhu2011}. However, these investigations all required low temperatures as the emitter dephasing occurs too rapidly or the energy dissipation is too high at ambient conditions.

The transition into the strong-coupling regime is dependent on the ability to control the electromagnetic coupling rate \(g\) between the emitter and the cavity. Since \( g \propto \mu_{\text{ex}} \sqrt{ N/V}\) \cite{Todisco2018}, where \(\mu_{\text{ex}}\) is the transition-dipole moment of the emitter, \(N\) is the number of excitons coherently coupled to the cavity, and \(V\) is the cavity mode volume, several key factors can be engineered to facilitate this transition. For purely dielectric cavities, the minimum mode volume is on the order of \((\lambda/n)^{3} \), thereby limiting the coupling strength achievable and requiring long coherence times in the cavity and emitter states to observe the strong-coupling regime. On the other hand, plasmonics offers the possibility for a significant reduction in the mode volume due to the sub-diffraction confinement of the electric field and the associated strong enhancement of the field strength at the price of higher losses, mainly due to absorption in the metal~\cite{Dezfouli2017}. Utilizing this to couple plasmons to various organic and inorganic semiconductors has gained~increasing attention in recent years~\cite{Torma2015,Marquier2017,Baranov2017a,Vasa2018,Goncalves2018,Fernandez-Dominguez2018}, and plasmonic systems have been shown to reach the strong-coupling regime when interacting with organic molecules~\cite{Berrier2011,Zengin2015,Chikkaraddy2016,Wersall2017,Todisco2018}, resulting in enhanced nonlinearities and low threshold polariton lasing and condensation~\cite{Ramezani2016,DeGiorgi2018}.

Regarding excitons in two-dimensional (2D) materials, mono- and multilayers of TMDCs have been the object of intense study in recent years in part due to their unique optical properties, capable of merging together the main properties of Frenkel excitons in organic semiconductors (high binding energy and large oscillator strength) with that of Wannier--Mott excitons in inorganic quantum wells (relatively large exciton radius, low photodegradation and small saturation density) at room temperature. However, the exciton-dipole moment is strongly oriented in the plane of the 2D material~\cite{Baranov2017a}, which requires special attention when coupling to external fields. Overcoming this challenge has resulted in studies demonstrating strong coupling of the excitons in 2D materials to plasmonic lattices~\cite{Liu2016,Wang2016a}, plasmonic and dielectric cavities~\cite{Liu2015c,Flatten2016,Hu2017,Kleemann2017,Han2018,Tserkezis2018}, plasmonic slit resonators~\cite{Gross2018}, as well as to individual plasmonic nanoparticles~\cite{Wen2017,Stuhrenberg2018}. Some studies have even reported strong coupling to charged excitons (trions) in WS\(_{2}\)~\cite{Cuadra2018}, and self-hybridization of the exciton with cavity modes in extended TMDC systems~\cite{Wang2016b,Yadgarov2018,Munkhbat2018}. The lattice structure of the TMDCs furthermore gives rise to a new degree of freedom in the form of valley polarization, which can be addressed using circularly polarized light when coupled to plasmonic structures~\cite{Gong2018,Chervy2018}.

In this work we present the observation of the strong-coupling regime between localized surface plasmons supported by chemically synthesized, single-crystalline gold nanodisks, and 2D excitons in mono- and multilayer WS2. Both systems exhibit strong in-plane optical response, and by careful control of the nanoparticles' size and position we achieve spectral and spatial overlap between the dipolar plasmon mode in the nanodisks and the A-exciton in WS\(_{2}\)~\cite{Cui2018,Gutierrez2013}. Since the nanodisks are sufficiently thin, the absorption dominates the extinction cross section, and we are therefore able to experimentally observe the onset of the strong-coupling regime in both scattering and reflection measurements with a Rabi splitting of \(\SI{\rabi \pm \rabiDelta}{meV}\) at room temperature. Inspired by recent findings~\cite{Stuhrenberg2018}, we furthermore present results obtained from multilayer WS\(_{2}\) resulting in strong coupling with a Rabi splitting of \(\SI{\rabiMulti \pm \rabiMultiDelta}{meV}\). To the best of our knowledge, this is the highest reported splitting in any TMDC coupled to an open plasmonic cavity. The accessibility of the cavity along with the clear presence of the strong-coupling regime is an important stepping stone towards further study of the plethora of fundamental physics in these strongly coupled light-matter systems, along with potential applications in quantum information processing, quantum metrology, nonlinear optical materials, and quantum light sources.

\section*{Results and discussion}

\begin{figure}
	\includegraphics{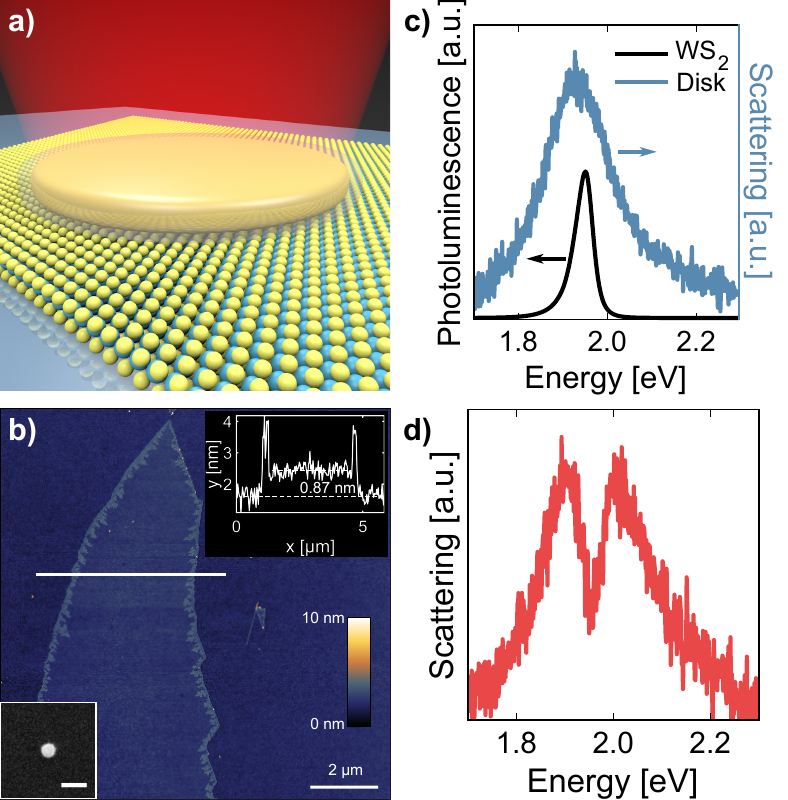}
    \caption{\textbf{(a)} Sketch of a single-crystalline gold nanodisk coupled to monolayer WS\(_{2}\). \textbf{(b)} Atomic force microscopy image of the monolayer WS\(_{2}\) flake. The inset in the lower left corner shows a scanning electron microscopy image of a typical nanodisk. Scale bar is \SI{100}{nm}. \textbf{(c)} Photoluminescence spectrum of monolayer WS\(_{2}\) (black) and dark-field scattering spectrum of an individual, uncoupled nanodisk (blue). \textbf{(d)} Experimental scattering spectrum of a single nanodisk coupled to WS\(_{2}\) clearly showing the splitting between the upper and lower polariton branch. \label{figure1}}
\end{figure}

In order to demonstrate the strong-coupling regime at room temperature, we deposit thin single-crystalline gold nanodisks on a monolayer of WS\(_{2}\) as shown in \autoref{figure1}a. The nanodisks were chosen for their strong in-plane, dipolar localized surface-plasmon resonances (LSPR) with narrow linewidths which are tunable by controlling their radius~\cite{Cui2018}. Like the other TMDCs, WS\(_{2}\) has a strong in-plane excitonic response with a large transition-dipole moment~\cite{Schuller2013,Baranov2017a}. Combined with the stability of the excitons under ambient conditions due to their large binding energies~\cite{Ramasubramaniam2012}, this makes WS\(_{2}\) highly suitable for exploring light-matter coupling at room temperature.

Flakes of WS\(_{2}\) were mechanically exfoliated from commercially available crystals (\href{http://www.hqgraphene.com/WS2.php}{HQ Graphene}) onto plasma cleaned SiO\(_{2}\)/Si substrates. Monolayer flakes were automatically detected and identified by quantitative optical mapping~\cite{Jessen2018}. The monolayer nature of these flakes was confirmed by atomic force microscopy (AFM) as depicted in \autoref{figure1}b, where the height of the flake was found to be \SI{\wsHeight}{nm} in agreement with previously reported results~\cite{Gutierrez2013}. The photoluminescence (PL) spectrum, black line in \autoref{figure1}c, reveals the exciton emission spectrum, where the asymmetry of the peak to the low-energy side is mainly caused by strong exciton-phonon coupling~\cite{Christiansen2017}, and also indicates the presence of the trion (charged exciton) in addition to the neutral state of the A-exciton~\cite{Currie2015}. Fitting a double Voigt profile to the PL spectrum yields emission energies of \(E_{\text{em,tri}} = \SI{\wExTrion}{eV}\) and \(E_{\text{em,ex}} = \SI{\wExNeutral}{eV}\), in accordance with the literature~\cite{Gutierrez2013} (see Figure S\(\suppPL\)). To extract the energy and linewidth of the \(A\)-exciton resonance, we performed reflection measurements on five different monolayer flakes, using spatially filtered white light to ensure normal incidence. Employing the transfer-matrix method\cite{Zhan2013} we obtained the dielectric function \(\varepsilon\) for the flakes from which we extract an average exciton energy \(E_{\text{ex}} = \SI{\wEx}{eV}\) and linewidth \(\gamma_{\text{ex}} = \SI{\gammaEx}{meV} \), see Figure S\(\suppReflFit\) and Table S\(\suppReflMonoPar\).

The open plasmonic cavities consist of circular single-crystalline gold nanodisks chemically synthesized as described previously~\cite{Cui2018}. Briefly, triangular gold nanoplates were first synthesized using a three-step seed-mediated method~\cite{Qin2016}. The nanodisks were produced by performing anisotropic oxidation on the nanoplates. The oxidation allows for finely tailoring the nanodisk diameter, and the thickness of the gold nanodisks remained unchanged during the oxidation process. A thin layer (\(\sim\)\SI{1}{nm}) of cetyltrimethylammonium bromide (CTAB) is present on the surface of the nanodisks after the synthesis~\cite{Cui2018}. In our study, circular gold nanodisks of \SI{8.8 \pm 0.2}{nm} in thickness, measured without the CTAB layer, were chosen because of their moderate plasmon damping and concomitant narrow plasmon linewidth. Due to the inherent variation in particle radii, each batch contains particles with their LSPR distributed around a mean energy, the latter chosen to match the exciton energy (\(E_{\text{ex}} = \SI{\wEx}{eV}\)). A representative scattering spectrum of a single gold nanodisk is shown in \autoref{figure1}c. Measuring the uncoupled nanodisks on the bare SiO\(_{2}\)/Si substrate results in an average linewidth of the plasmon resonance of \(\gamma_{\text{pl}} = \SI{\gammaSp}{meV}\). The variation of the particles' LSPRs is exemplified in ensemble measurements performed on the nanodisk solution in Figure S\(\suppEnsemble\),
where the extinction spectra show a width about twice that of the individual nanodisks due to inhomogeneous broadening. With an energy of the resonance around \SI{1.96}{eV} we obtain a \(Q\)-factor of \( E_{\text{pl}} / \gamma_{\text{pl}} \approx 11.5 \) close to its intrinsic quasistatic limit of \(\sim\)\(\QMax\) for an arbitrarily shaped gold nanoparticle when only the dielectric function is considered~\cite{Wang2006}. This shows that other dissipation pathways, such as surface and imperfection scattering, are minimized due to the single-crystalline nature of the nanodisks.

\begin{figure*}
	\includegraphics{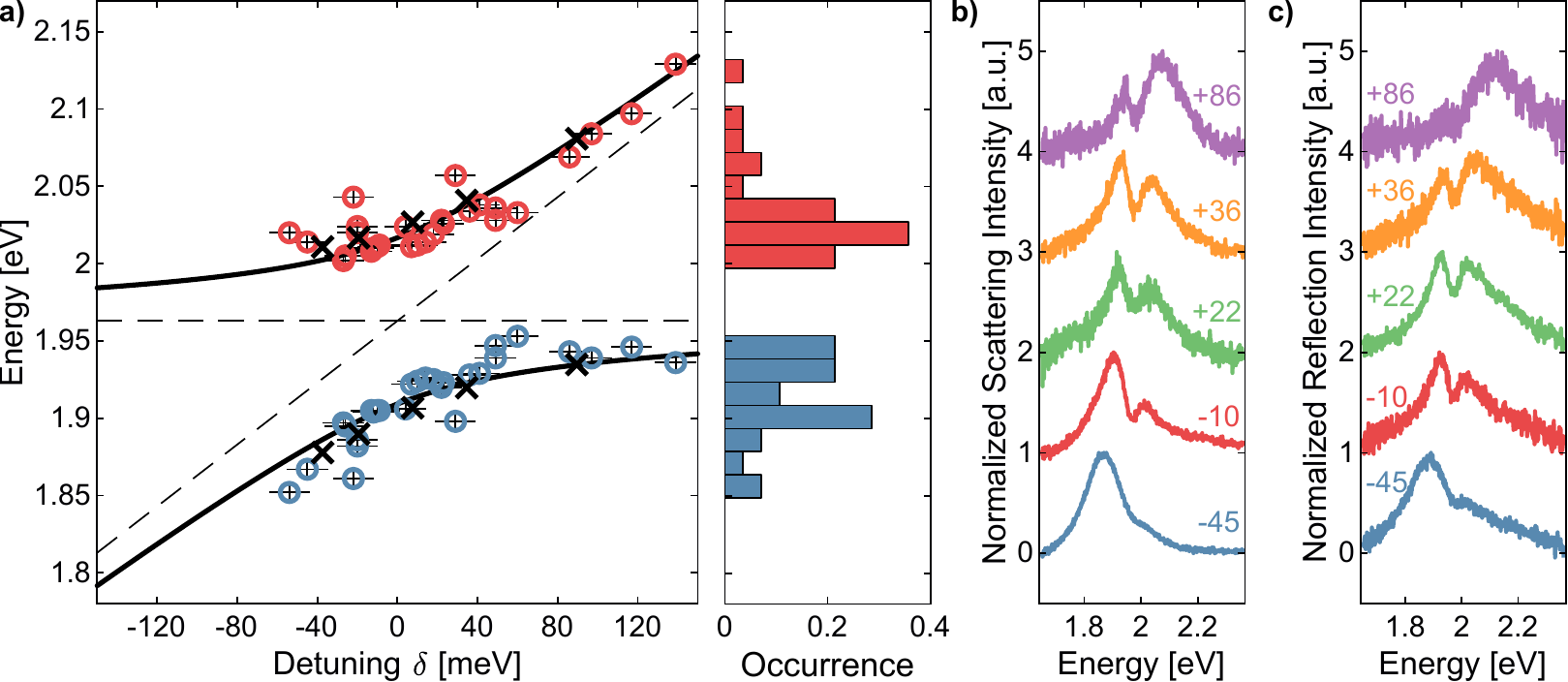}
    \caption{\textbf{(a)} Experimental dispersion obtained from scattering measurements for the coupled plasmon-exciton system on monolayer WS\(_{2}\) with the lower (blue) and upper (red) polariton branch. The right panel shows the distribution of measured peaks. The solid black line is a fit to the data using \autoref{eq:eigenvalues}, while the horizontal and sloped dashed lines indicate the exciton and plasmon energies, respectively. The black crosses are the peak positions extracted from the calculated spectra shown in \autoref{figure3}b. \textbf{(b)} Experimental dark-field scattering spectra from five different single gold nanodisks showing the evolution of the optical response with different plasmon energies. The individual detunings \( \delta = E_{\text{pl}} - E_{\text{ex}} \) are presented next to the spectra. \textbf{(c)} Experimental reflection spectra for the same five particles as in \textbf{(b)}. The absorption of the disks dominates the extinction spectra by approximately an order of magnitude making the reflection spectrum representative of the particle absorption (see \autoref{figure3}a). \label{figure2}}
\end{figure*}

Achieving the strong electromagnetic coupling requires simultaneous spatial and spectral overlap of the LSPR and the exciton in the nanodisk and WS\(_{2}\), respectively. The spatial overlap was ensured by dropcasting the nanodisks directly onto the WS\(_{2}\) flakes. The CTAB molecule layer ensures a narrow gap without direct contact between the metal and the underlying flake in order to prevent possible quenching. The presence of a high refractive-index material under the particle furthermore serves to concentrate the electric field in the WS\(_{2}\) and in the substrate rather than in the surrounding air.
Spectral overlap was achieved by precise control of the nanodisk radius. We took advantage of the aforementioned natural variation of the nanodisk radii inherent in a batch, as this allowed for measurements on different gold nanodisks with their plasmon energies distributed around the energy of the exciton.
The positions of the particles were then examined by dark-field (DF) microscopy, and scattering spectra were obtained for each individual particle. After the measurements, scanning electron microscope images were taken to ensure the measured spectra originate only from single nanodisks rather than from dimers or individual particles of more complex shape. A spectrum taken from a nanoparticle resting on a WS\(_{2}\) monolayer is shown in \autoref{figure1}d, where two distinct peaks appear around the bare exciton energy. The almost equal strength of the two peaks indicates that the LSPR of the nanodisk is close to the exciton resonance. By measuring particles with different radii, and thus different LSPR energies, we are in this manner able to map out the dispersion of the coupled system as a function of the plasmon-exciton energy detuning. All the examined spectra show a double peaked lineshape. Extracting the energy of the peaks from the scattering spectra, a clear anti-crossing behaviour is observed as shown in \autoref{figure2}a, which is characteristic for these emitter-cavity coupled systems~\cite{Chikkaraddy2016,Wersall2017,Yoshie2004}.

In order to analyze the data, we apply the coupled-oscillator model in its non-Hermitian Hamiltonian form given by the eigenvalue problem~\cite{Flick2018}:
\begin{equation} \label{eq:couplingHamiltonian}
\left[ \begin{matrix}
E_{\text{pl}} - i \frac{\gamma_{\text{pl}}}{2} & g \\
g & E_{\text{ex}} - i \frac{\gamma_{\text{ex}}}{2}
\end{matrix} \right] \left[ \begin{matrix}
\alpha \\ \beta
\end{matrix} \right]_{\pm} = E_{\pm} \left[ \begin{matrix}
\alpha \\ \beta
\end{matrix} \right]_{\pm} ,
\end{equation}
where \(E_{\text{pl}}\) and \(\gamma_{\text{pl}}\) are the plasmon energy and linewidth, \(E_{\text{ex}}\) and \(\gamma_{\text{ex}}\) are the exciton energy and linewidth, \(g\) is the coupling strength, and \(\alpha\) and \(\beta\) are the Hopfield coefficients describing the polaritonic state in terms of its plasmonic (\(\alpha_{\pm}\)) and excitonic (\(\beta_{\pm}\)) constituents. Diagonalizing the Hamiltonian yields the two energy eigenvalues
\begin{align}
E_{\pm} = \frac{1}{2} &\left(E_{\text{pl}} + E_{\text{ex}}\right) - \frac{i}{4}\left(\gamma_{\text{pl}} + \gamma_{\text{ex}}\right) \nonumber \\
&\pm \sqrt{g^{2} + \frac{1}{4} \left( \delta - \frac{i}{2} \left(\gamma_{\text{pl}} - \gamma_{\text{ex}} \right) \right)^{2}}, \label{eq:eigenvalues}
\end{align}
where we have introduced the detuning \(\delta = E_{\text{pl}} - E_{\text{ex}} \) of the plasmon with respect to the exciton energy. The eigenvalues describe an energy spectrum divided into the upper polariton branch (UPB, \(E_{+}\)) and the lower polariton branch (LPB, \(E_{-}\)). The vacuum Rabi splitting \(E_{\text{Rabi}}\) is then defined as the energy difference between the UPB and LPB exactly at zero detuning, giving
\begin{equation} \label{eq:rabiSplitting}
E_{\text{Rabi}} = \sqrt{4 g^{2} - \frac{\left( \gamma_{\text{pl}} - \gamma_{\text{ex}} \right)^{2}}{4}}.
\end{equation}
From the scattering spectra of the individual nanodisks on WS\(_{2}\) we read off the respective energies of the upper and lower polariton branches, \(E_{+}\) and \(E_{-}\). From energy conservation we then obtain the plasmon energy as \(E_{\text{pl}}~=~E_{+}~+~E_{-}~-~E_{\text{ex}} \). The avoided crossing appears clearly on the histogram in the right side of \autoref{figure2}a, with no data points located between the two polaritonic branches. The vertical error bars account for a read-off uncertainty of the energies \(E_{\pm}\) which we estimate to be \(\pm \SI{\readoffDelta}{\meV}\). The horizontal error bars represent the propagated uncertainties in the calculated detunings, which are dominated by the standard deviation of the measured exciton energy. 

To estimate the coupling strength \(g\) and the vacuum Rabi splitting \(E_{\text{Rabi}}\), we simultaneously fit the real part of the eigenvalue spectrum \autoref{eq:eigenvalues} to the experimentally determined dispersion of the UPB and LPB. In the fitting procedure we fix the values of the exciton energy, (\( E_{\text{ex}} = \SI{\wEx}{eV}\)), the exciton linewidth (\( \gamma_{\text{ex}} = \SI{\gammaEx}{meV}\)), and the plasmon linewidth (\( \gamma_{\text{pl}} = \SI{\gammaSp}{meV}\)) to the experimentally determined values, which leaves the coupling strength as the only free parameter. This procedure yields a value of \(g = \SI{\coupling \pm \couplingDelta}{meV}\), from which we calculate a Rabi splitting of \(E_{\text{Rabi}} = \SI{\rabi \pm \rabiDelta}{meV} \) using \autoref{eq:rabiSplitting}. Comparing this with the overall losses in the system \(\frac{1}{2}\left(\gamma_{\text{pl}} + \gamma_{\text{ex}}\right) = \SI{\gammaAvg \pm \gammaAvgDelta}{meV} \), we see that our hybrid nanodisk/WS\(_{2}\) system is at the onset of the strong-coupling regime since as per the criterion $E_{\text{Rabi}} > \frac{1}{2}\left(\gamma_{\text{pl}}+ \gamma_{\text{ex}}\right)$.

Five scattering spectra showing the evolution of the optical response of the hybrid system are depicted in \autoref{figure2}b. Here we clearly see the avoided crossing in the succession of individual spectra, where the scattering amplitude is shifting from the LPB to the UPB as the detuning is scanned across the direct energy match between plasmon and exciton at \(\delta = \SI{0}{meV}\). In addition, we note that the noise generally increases for larger \(\delta\) since the LSPR resonance energy increases with decreasing particle radius, resulting in a lower scattering cross section.

Since \( \delta = E_{+} + E_{-} - 2 E_{\text{ex}} \), the exact value for the detuning reported in \autoref{figure2}b-c is highly sensitive to the value of \(E_{\text{ex}}\). From reflection measurements we observe exciton energies in the range \SIrange{1.958}{1.968}{eV}, that can result as a consequence of local strain or doping from the nanoparticle solution during deposition. Extracting the exciton energy from the fitting with the coupled oscillator model we obtain a slightly higher value (E-ex=1.981+-0.04 eV, see Figure S5) that fits within the statistical error from the fitting procedure and does not affect the evaluated Rabi splitting (see Figure S6).

Since \( \delta = E_{+} + E_{-} - 2 E_{\text{ex}} \), the exact value for \(\delta\) reported in \autoref{figure2}b-c is highly sensitive to the value of \(E_{\text{ex}}\). From the reflection measurements we observe exciton energies in the range \SIrange{1.958}{1.968}{eV} giving a detuning variation of \(\Delta \delta = \SI{20}{meV} \), which could be expected to influence the resulting fit as well. To validate our results, we therefore also use the coupled-oscillator model to extract the peak positions of the UPB and LPB, and to obtain the exciton energy~\cite{Wu2010a}, see Figure S\suppComSpec. From this we find a slightly higher exciton energy of \(E_{\text{ex}} = \SI{1.981}{eV}\), but also a larger spread of \SI{40}{meV} (i.e. $\Delta \delta = \SI{80}{meV}$). This larger variation and change in \(E_{\text{ex}}\) can be caused by e.g. local strain or doping from the nanoparticle solution~\cite{Cong2018}. However, using instead this average in the same analysis as before results in a change in the Rabi splitting of only \SI{4}{meV}, despite the shift of \SI{36}{meV} in the detuning for all measurement points, see Figure S\(\suppAvoidCrossCom\). In any case we can conclude from this analysis that our system is at the onset of the strong-coupling regime whether we use one or the other of the above methods to determine \(E_{\text{ex}}\).

\begin{figure}
	\includegraphics{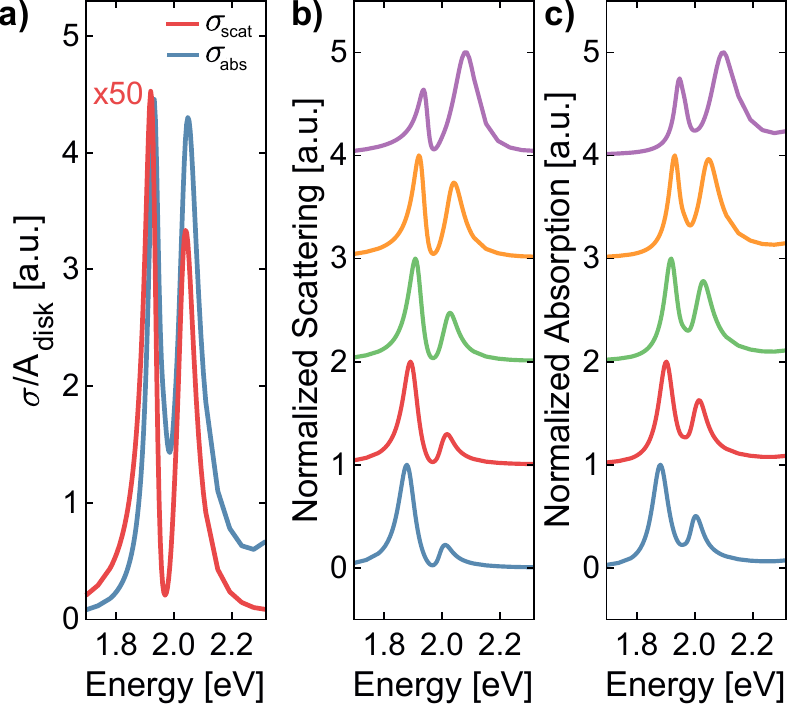}
    \caption{\textbf{(a)} Calculated scattering (red) and absorption (blue) cross sections \(\sigma\) obtained with finite-element method calculations of a nanodisk of radius \(R = \SI{\rTwo}{nm}\) normalized to the geometrical area. The scattering cross section has been multiplied by 50 to fit the scale. The ratios between the maxima of \(\sigma_{\text{abs}}\) and \(\sigma_{\text{scat}}\) vary from 83 (\(R = \SI{\rOne}{nm}\)) down to 17 (\(R = \SI{\rFive}{nm}\)). \textbf{(b-c)} Normalized scattering and absorption of particles of radii \(R = \SI{\rOne}{nm}\), \SI{\rTwo}{nm}, \SI{\rThree}{nm}, \SI{\rFour}{nm}, and \SI{\rFive}{nm}, arranged from top to bottom.\label{figure3}}
\end{figure}

The strong-coupling regime is associated with a pronounced mode splitting not only in the scattering but also in the absorption spectrum of the coupled system~\cite{Antosiewicz2014}. Usually, the presence of this splitting in absorption is only verified numerically~\cite{Wen2017,Stuhrenberg2018}, since measuring the absorption independently from the scattering  normally requires specialized techniques~\cite{Yorulmaz2015}. However, as our nanodisks have small radii (\(<\!\SI{35}{nm}\)) and a thickness of only \SI{8.8}{nm}, our numerical calculations show that the absorption cross section is larger than the scattering cross section by more than an order of magnitude, as depicted in \autoref{figure3}a, and thus dominates the reflection spectrum entirely. By measuring the reflection spectra of the particles, we therefore have direct access to the absorption spectrum and through that experimental verification of the coupling regime. In \autoref{figure2}c the normalized reflection spectra $\Delta \mathcal{R}$ of the same five particles as in \autoref{figure2}b are shown. The spectrum has been calculated as
\begin{equation}
\Delta \mathcal{R} = \frac{\mathcal{R}_{\text{WS\(_{2}\)}} - \mathcal{R}_{\text{disk}}}{\mathcal{R}_{\text{sub}}},
\end{equation}
where \(\mathcal{R}_{\text{WS$_{2}$}}\), \(\mathcal{R}_{\text{disk}}\), and \(\mathcal{R}_{\text{sub}}\) are the light intensities reflected from the WS\(_{2}\), from the nanodisk on the WS\(_{2}\), and from the bare SiO\(_{2}\)/Si substrate, respectively. As \(E_{\text{pl}}\) is tuned across the exciton energy, we observe the same behaviour for the reflection as for the scattering spectra, but we note that the particle absorption is blueshifted in all cases compared to the scattering, which is also observed in our calculations, see \autoref{figure3}b-c. This shift is predicted from Mie theory and has also been demonstrated experimentally~\cite{Yorulmaz2015}. The measured mode splitting also in the reflection spectra corroborates the observation of the onset of the strong-coupling regime in our system.

To further support the experimental observations, we perform finite-element method (FEM) calculations of the electrodynamics (see methods part). We use the experimentally determined radii (from \SIrange{\rOne}{\rFive}{nm})\cite{Cui2018} and nanodisk thickness (\SI{9}{nm}), as well as a \SI{1}{nm} thick polymer layer surrounding the particle. We furthermore use our experimentally determined dielectric function of the WS\(_{2}\) flakes. For the gold, we use the values for \(\varepsilon\) measured by~\citet{McPeak2015a} in optimized film quality conditions. This choice of reference data was made to better reflect the single-crystalline nature of our nanoparticles. In this way, we obtain scattering and absorption spectra of the particles as shown in \autoref{figure3}b and c. Since the nanodisks are thinner than the skin depth of gold of around \( \SI{25}{nm}\) at \SI{1.96}{eV}~\cite{Olmon2012}, the electric field completely permeates the particles, making the results very sensitive to the particular value of the dielectric function. For instance, using instead the values from Johnson and Christy~\cite{Johnson1972} redshifts the resonance position by around \SI{40}{meV} (see Figure S\suppNorJC), and changes in especially the imaginary part of \(\varepsilon\) have been reported for thin gold films~\cite{Arsenin2017}. However, we still see that we are able to reproduce the experimentally observed behaviour both in scattering and absorption with good agreement, which further corroborates the observation of the onset of the strong-coupling regime in our system. The calculations show cross sections approaching zero at the exciton energy, indicative of strong plasmon-exciton coupling~\cite{Antosiewicz2014}. These same near-zero intensities are not observed in the experiments, which can be caused by a signal background coming from the strong scattering of the WS\(_{2}\) flake edges. Although we perform background removal on all measured spectra, the scattering cross section of the nanodisks is small and any remaining background is expected to influence the overall signal intensity.

Extracting the peak locations of the calculations in the same manner as for the experimental data allows us to compare the position of the UPB and LPB to the experimental results, as indicated by the black crosses in \autoref{figure2}a, where we see excellent agreement between experiments and calculations.

\begin{figure}
	\includegraphics{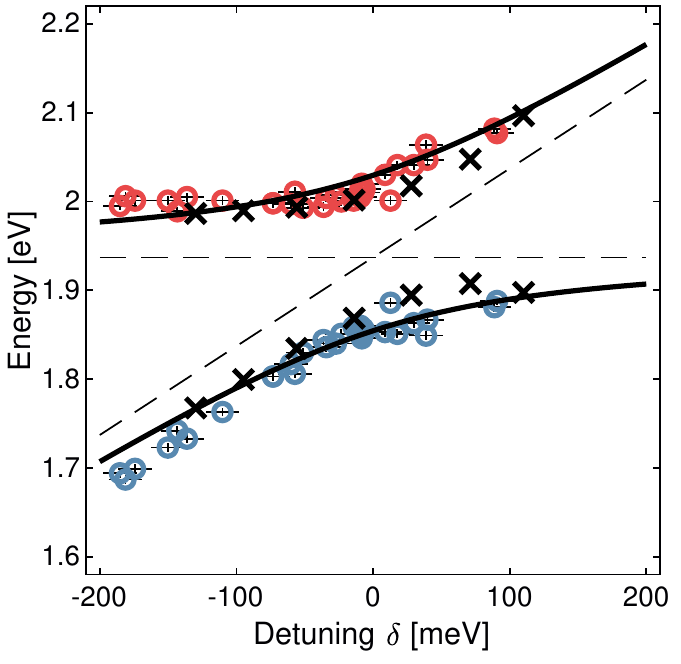}
    \caption{Experimentally obtained dispersion for the coupled plasmon-exciton system on multilayer WS\(_{2}\). Note the change in scale of the \(y\)-axis compared to \autoref{figure2}a. \label{figure4}}
\end{figure}

Encouraged by the observation of strong coupling of nanodisks on a monolayer, we then went on to study how these gold disks coupled to WS\(_{2}\) multilayers. Earlier experiments by~\citet{Kleemann2017} and by~\citet{Stuhrenberg2018} have shown the possibility to increase the Rabi~splitting by increasing the number of TMDC layers coupled to the plasmonic particle. In the former case the effect comes from a higher out-of-plane excitonic dipole moment in a particle on a mirror cavity, while in the latter case the increase is caused both by a lowering of the mode volume and an increase in the total mode overlap with the excitonic material. A significant increase in the Rabi splitting of \(\sim \! \SI{22}{\%}\) was observed by St{\"u}hrenberg and coworkers when going from one to four layers of WSe\(_{2}\), after which the effect saturates. In order to asses the magnitude of this effect in our system, we measure the optical response of particles on a WS\(_{2}\) flake \SI{4.4}{nm} in height corresponding to seven layers. Using the transfer-matrix method again allows us to extract the dielectric function, giving \(E_{\text{ex,multi}} = \SI{\wExMulti}{eV}\) and \(\gamma_{\text{ex,multi}} = \SI{\gammaExMulti}{meV} \) (see Table S\(\suppReflMultiPar\)), where both the broadening and redshift of the A-exciton resonance are known phenomena for multilayers~\cite{Stuhrenberg2018}. Following the same procedure for data analysis as for the monolayer, we obtain the experimental results plotted in \autoref{figure4}, where we again see the avoided crossing characteristic for these systems. In accordance with the previous results, we also find an increase in the coupling strength to \(g_{\text{multi}} = \SI{\couplingMulti \pm \couplingMultiDelta}{meV}\) (an increase of \SI{\gIncrease}{\%}), resulting in a Rabi splitting of \(E_{\text{Rabi,multi}} = \SI{\rabiMulti \pm \rabiMultiDelta}{meV} \) (increased by \SI{\rabiIncrease}{\%}). To the best of our knowledge, this is the highest Rabi splitting reported in a TMDC coupled to an open plasmonic cavity \cite{Han2018, Wen2017, Zheng2017b, Stuhrenberg2018}. Even larger splittings can be obtained using the 2D material itself as both emitter host and cavity~\cite{Wang2016b,Munkhbat2018,Yadgarov2018}, or by coupling TMDCs to both cavity and localized plasmon modes simultaneously \cite{Bisht2018}. However, the direct availability of the open plasmonic cavity is advantageous for future near-field explorations and sensing applications, as well external coupling to other cavities and optical circuit elements.

In order to understand the important increase in the coupling strength that we observe for the seven-layer WS\(_{2}\) flake, we perform numerical calculations using the dielectric function extracted from our reflection measurements. We again extract the position of the UPB and LPB as shown by the black crosses in \autoref{figure4}. The calculations predict an overall increase of the coupling strength with the number of layers. However, quantitatively we observe a slightly lower increase in the Rabi splitting in comparison to our experimental data. We ascribe this difference to the fact that we have approximated the optical response of multilayer WS\(_{2}\) as an isotropic dielectric function and we therefore do not take into account the out-of-plane response of multilayer WS\(_{2}\). Despite these quantitative differences we observe qualitative agreement between our calculations and our experimental data. We use our numerical model to estimate the change in the effective mode volume $V_{\rm eff}$ (see \autoref{eq:modeVolume} in Methods) when increasing the number of layers from one to seven. For the monolayer we estimate $V_{\rm eff}$ to be \SI{900}{nm^{3}}, while for the multilayer this number naturally increases to \SI{2100}{nm^{3}} due to the seven times larger integration volume $\cal V$. This larger effective mode volumes means that a larger number of excitons can couple coherently with the plasmon field, thus increasing the overall coupling strength. Note however, that relatively the effective mode volume calculated for each layer decreases significantly to an average value of \SI{390}{nm^{3}}. This is caused by a higher permittivity of the environment experienced by each of the seven layers. A higher number of excitons involved in the coupling can explain the substantial change in the Rabi splitting observed experimentally between mono- and multilayers WS\(_{2}\). This conclusion is in agreement with recent experimental works~\cite{Stuhrenberg2018,Kleemann2017}.  

\section*{Conclusion}
We have successfully demonstrated the onset of the strong-coupling regime between monolayer WS\(_{2}\) flakes and single-crystalline gold nanodisks with a Rabi splitting of \SI{\rabi \pm \rabiDelta}{meV}. The natural variation in the nanodisk radius allowed us to probe the coupled system around the exciton energy and map out the avoided crossing.
Spurred by recent experiments, we furthermore demonstrated an even stronger coupling between the nanodisks and multilayer WS\(_{2}\) yielding a Rabi splitting of \SI{\rabiMulti \pm \rabiMultiDelta}{meV}, i.e. in excess of seven times $k_BT$ at room temperature. This is enabled by a larger overlap of the plasmon mode and the increased number of excitons located in multilayer WS\(_{2}\). This way of controlling the coupling regime in plasmonic nanodisks and structurally similar systems is a key component towards applications within all-optical circuitry, polaritonic lasers, and quantum information processing, as well as the exploration of possible anti-bunching effects in multi-emitters coupled to plasmonic cavities~\cite{Saez-Blazquez2017,Fernandez-Dominguez2018}. 

\section*{Methods}

\subsection*{Optical measurements}
A custom spectroscopy setup built from a Nikon Eclipse Ti-U inverted microscope was used for the optical measurements. For the dark-field and bright-field spectra a halogen lamp with a tunable power up to \SI{100}{W} was used, while for the photoluminescence a \SI{407}{nm} diode laser (Integrated Optics) was used. The light was focused on the sample with a TU Plan Fluor objective from Nikon (100$\times$, \SI{0.9}{NA}) and collected with the same objective. Afterwards, the light was directed toward a slit allowing for precise selection of the collection area. The light then entered a Shamrock 303i Spectrometer equipped with a \SI{450}{nm} longpass filter (FELH0450 from Thorlabs) and an electronically cooled Newton 970 EMCCD for acquiring spectra. All spectra consist of the sum of several lines on the 2D CCD uniquely identified for each particle. Each of these spectra were first corrected for dark counts, and then for background using the local environment in close vicinity of the individual particles obtained directly from the same spectroscopic image. Finally, the spectra were divided by the normalized white light spectrum of the halogen lamp. The photoluminescence spectra were obtained in a similar manner apart from the white light spectrum correction. All experiments were performed at room temperature.

\subsection*{Finite-element calculations}

A commercially available FEM software (COMSOL Multiphysics v. 5.3a) was used to simulate the electrodynamic response of the nanodisks placed on top of mono- and multilayer WS\(_{2}\). We used experimental measurements of the nanoparticles to estimate their radii, which gave us nanodisks with radii ranging from \SI{19}{nm} to \SI{29}{nm} for our calculations on monolayer WS\(_{2}\). Due to higher values of the dielectric function and a higher thickness of the multilayers WS\(_{2}\), the LSPR of the nanodisks were redshifted and the radii close to the excitonic resonance ranged from \SI{11}{nm} to \SI{21}{nm}. In all our calculations, we have used a thickness of \SI{9}{nm} for the metallic core~\cite{Cui2018}. For this we used the dielectric function measured by~\citet{McPeak2015a} to better reflect the single-crystalline nature of our nanoparticles.
We model the layer of CTAB molecule as a homogeneous \SI{1}{nm} thick dielectric layer surrounding the nanodisks with a refractive index of 1.435~\cite{Kekicheff1994}. The nanodisk with the CTAB layer is in direct contact with WS\(_{2}\). The one and seven layers WS\(_{2}\) are modelled as homogeneous, isotropic bulk materials with thicknesses of \SI{0.6}{nm} and \SI{4.3}{nm}, respectively, and the dielectric functions for both cases are extracted from experimental reflection measurements followed by a treatment with the transfer-matrix method (see Table S1 and S2). The SiO\(_{2}\) substrate is modelled as a semi-infinite layer and the dielectric function is taken from Malitson~\cite{Malitson1965a}. The exciting field is impinging at normal incidence onto the nanodisk. The scattering cross section is evaluated by integrating the scattered energy density flow through a surface with a radius of \SI{100}{nm} placed \SI{150}{nm} above the nanodisk to simulate the numerical aperture of the objective used in our measurements (see optical measurements section above). The absorption cross section is calculated by integrating the energy density dissipated within the volume of the metallic core of the nanodisk and within the layered material WS\(_{2}\) located below the nanodisk.
The mesh size inside the WS\(_{2}\) layers, the CTAB layer and in the metal core close to the metal/CTAB interface is as small as \SI{0.3}{nm} to allow numerical convergence. The rest of the core of the nanodisk is meshed with elements with sizes ranging from \SI{0.3}{nm} to \SI{4}{nm}. 
For the effective mode volume calculations (see section below), we calculate \autoref{eq:modeVolume} within a volume $\cal V$ inside WS\(_{2}\) (taking into account the spectral dispersion) and restricted within a radius of 50\,nm around the nanodisks. In the special case of the seven-layer WS\(_{2}\), we calculate the effective mode volume $V_{\rm eff}$ for each of the seven layers.

\subsection*{Effective mode volume calculations}

In cavity quantum-electrodynamics (QED), the mode volume $V$ associated with a single exciton in the point ${\bf r}_0$ is commonly given by~\cite{Koenderink2010}

\begin{equation}
V=\frac{\sum_{\alpha=x,y,z}\int d{\bf r}\,\epsilon({\bf r})| E_\alpha ({\bf r})|^2}{\sum_{\alpha=x,y,z}\epsilon ({\bf r}_0) | E_\alpha({\bf r}_0)|^2},
\end{equation}
where the electrical field $\bf E$ is evaluated at the exciton frequency, and with the integral over the energy density extending over all space. The energy density can be appropriately corrected to account for dispersive media, such as metallic elements~\cite{Koenderink2010}. The lower part of the fraction contains the field enhancement at the position of the exciton, including an implicit average over all possible dipole directions.

To account for multiple excitons distributed in a quasi 2D layer, we calculate the effective mode volume $V_{\rm eff}$ following the prescription by~\citet{Wen2017} as

\begin{equation}\label{eq:modeVolume}
 V_{\rm eff}=\frac{\sum_{\alpha=x,y}\int_{\cal V} d{\bf r}\,\epsilon({\bf r})| E_\alpha ({\bf r})|^2}{\sum_{\alpha=x,y}\epsilon ({\bf r}_0) | E_\alpha({\bf r}_0)|^2},
\end{equation}
where ${\bf r}_0$ is now located within $\cal V$, and $\epsilon$ is the dielectric function of the layered 2D material (in the absence of the exciton transition). Perhaps intuitively, the spatial integral over the energy density is restricted to the volume $\cal V$ of the layered 2D material, including only in-plane field components $E_x$ and $E_y$ in accordance with the in-plane orientation of the dipole moments of the 2D excitons.
We emphasize that this expression can be substantiated through rigorous summation of spatially uniformly distributed dipoles within the volume $\cal V$ of the layered 2D material, while subsequently accommodating an effective number ($N_{\rm eff}$) of dipoles in the point ${\bf r}_0$, eventually causing the same effective coupling constant $g\propto \sqrt{N_{\rm eff}/V_{\rm eff}}$. As a result, $N_{\rm eff} = \rho \times V_{\rm eff}$, where $\rho$ is the volume density of dipoles throughout the homogeneous layered 2D material.

\section*{Acknowledgments}

We thank C. Tserkezis and J. M. Hvam for stimulating discussions.
The Center for Nanostructured Graphene is sponsored by the Danish National Research Foundation (Project No. DNRF103).
F.~T. acknowledges a MULTIPLY fellowship under the Marie Sk\l{}odowska-Curie COFUND Action (Grant Agreement No. 713694)
N.~A.~M. is a VILLUM Investigator supported by VILLUM FONDEN (Grant No. 16498).
T.~R. and A.~B. acknowledge the IDUN Center of Excellence funded by the Danish National Research Foundation (Project No. DNRF122) and VILLUM FONDEN (Grant No. 9301).

\providecommand{\latin}[1]{#1}
\makeatletter
\providecommand{\doi}
  {\begingroup\let\do\@makeother\dospecials
  \catcode`\{=1 \catcode`\}=2\doi@aux}
\providecommand{\doi@aux}[1]{\endgroup\texttt{#1}}
\makeatother
\providecommand*\mcitethebibliography{\thebibliography}
\csname @ifundefined\endcsname{endmcitethebibliography}
  {\let\endmcitethebibliography\endthebibliography}{}

\end{document}